\documentclass[10pt]{article}

\begin{document}

\title{\Large \bf On the nature of "hidden symmetry" or "accidental degeneracy" in the
Kepler problem
\author{\large Tamar~T.~Khachidze$^*$, Anzor~A.~Khelashvili$^*$ \bigskip \\
{\it  $^*$~ Department of Theoretical Physics, Ivane Javakhishvili
 Tbilisi State University,}
\\ {\it I.Chavchavadze ave. 3, 0128, Tbilisi, Georgia} \\ } }

\maketitle

 {\large
 Lecture given at the \textit {Summer School and Conference "New
 Trends of High Energy Physics", Yalta, Crimea, Ukraine,16-23 Sept.
 2006}
\\

{\bf Abstract}\\
\medskip
 We propose a symmetry of the Dirac equation
under the interchange of signs of eigenvalues of the Dirac's $K$
operator. We show that the only potential which obeys this
requirement is the Coulomb one for both vector and scalar cases.

\section*{} \label{s1}

The problem is rather old, but several principal questions remain
without response. The Kepler problem has an additional symmetry,
more precisely, additional conserved quantity, so-called
Laplace-Runge-Lenz (LRL) vector ~\cite{Connel}.  Because of this
fact there appears extended algebra together with orbital
momentum, which is isomorphic to $SO(4)$ (for negative total
energies). But components of LRL vector as generators of this
algebra do not participate in any geometric transformations. In
1926 W.Pauli~\cite{Pauli} considered this algebra in quantum
mechanics and obtained Hydrogen atom spectrum by only algebraic
methods. In 30-ies V.Fock~\cite{Fock} considered the Schroedinger
equation for the Kepler problem in momentum representation and had
shown that the Coulomb spectrum has an $SO(4)$ symmetry in
energy-momentum space. Afterwards algebraic methods attract more
wide applications and it was cleared up in the last decades that
the hidden symmetry of the Kepler problem is closely related to
the supersymmetry of Hydrogen atom.

   Because of its significance this symmetry appears even in the textbooks.
   For example, in Classical Mechanics by
   J.Goldstein~\cite{Goldstein}, as well as in recent books of
   Quantum Mechanics~\cite{Bohm}. As regards of relativistic
   quantum mechanics (Dirac equation) relatively less is known,
   although the so-called radial supersymmetry was demonstrated a
   long-tome ago~\cite{Sukumar} and the many problems are studied
   rather well.

 The aim of our talk is to determine what the Dirac equation tells
 us about this problem. We can see that very interesting physical
 picture emerges.

   For the Kepler problem in the Dirac equation Johnson and
   Lippmann ~\cite{Johnson} published very brief abstract in
   Physical Review at 1950. They had written that there is an
   additional conserved quantity
\begin{equation}
A=\vec{\sigma}\cdot \vec{r}r^{-1} - \imath (\frac{\hbar
c}{e^2})(mc^2)^{-1}j\rho_{1}(H-mc^2\rho_3),
\end{equation}
 which plays the same role in Dirac equation, as the LRL vector
  in Schroedinger equation.

   As regards to the more detailed derivation,
 to the best our knowledge,  is not published in scientific literature
 (one of the curious fact in the history of physics of 20th
century)
      Moreover as far as commutativity    of the Johnson-Lippmann(JL) operator
       with the Dirac Hamiltonian is concerned, it is usually mentioned
      that it can be proved by  "rather tedious  calculations"~\cite{Katsura}.

     Recently~\cite{Khachidze,Khachidze-2}, we developed rather
     simple and transparent way for deriving the JL operator. We
     obtained this operator and at the same time proved its
     commutativity with the Hamiltonian.

      After we consider necessity to be convinced, that the Coulomb problem
      is distinguishable in this point of view. We considered the
      Dirac equation in arbitrary central potential, $V(r)$ and
      shown that the symmetry, which will be defined more precisely
      below, takes place only for Coulomb potential.

       We have chosen the subject of today's talk so that it is
       consonant with 1966 School,40th Years of which today's School is devoted to,
        when several lectures were dedicated to the problem of hidden symmetry - lectures by
       Alilluev, Matveenko, Popov and Werle~\cite{Proceedings}.
       During several decades these lectures were the only source,
       from which the young students and scientists could learn about
       this topic and the Proceedings of that School is
       an excellent monograph in many subjects of Theoretical High
       Energy Physics.

        In some respects our today's lecture is a continuation
        (but not analytic) of those old lectures.

            First of all consider the Dirac Hamiltonian in central
            field, $V(r)$, which is the 4th component of the
            Lorentz vector:

\begin{equation}
H = \vec{\alpha}\cdot \vec{p}+\beta m +V(r)
\end{equation}

            The so-called Dirac's $K$-operator, defined as
            \begin{equation}
K = \beta (\vec{\Sigma} \cdot \vec{l} +1),
\end{equation}
commutes with this Hamiltonian for arbitrary $V(r)$,  $[K,H]=0$.
Here $\vec{\Sigma}$ is the spin matrix
 $$
 \vec {\Sigma}  = \left( \begin{array}{l}
 \vec {\sigma} \ 0 \\
 0 \ \vec {\sigma}  \\
 \end{array} \right) \
$$
            and    $\vec{l}$    - orbital momentum vector-operator.

               It is evident that the eigenvalues of Hamiltonian
               also are labeled by eigenvalues of $K$. For
               example, the well-known \textbf{Sommerfeld formula} for the
               Coulombic spectrum looks like
$$ E_{n,|k|}= m \left[
1+\frac{(Z\alpha)^2}{(n-|\kappa|+\sqrt{\kappa^2-(Z \alpha)^2})^2}
\right]^{-1/2} $$

               It manifests explicit dependence on $\kappa$  , more
               precisely on $|\kappa|=j+1/2$.
                  It is surprising that for other solvable
                  potentials the degeneracy with respect to signs
                  of     $\kappa$   does not take place.

                         \textbf{Is it peculiar only for Coulomb
                          potential?}

              Let us consider Hamiltonian (2) with arbitrary
              potential and \textbf{require the degeneracy in} $\pm \kappa$.
              It is natural that for description of this
              degeneracy one has to find an operator, that mixes these
              two signs. It is clear that such an operator, say
              $A$, must be anticommuting with $K$, i.e.
\begin{equation}
\{A, K\} = AK+KA = 0
\end{equation}

              If at the same time this operator  commutes
              with the Hamiltonian, then it'll generate the
              symmetry of the Dirac equation.
                Therefore, we need an operator $A$ with the
                following properties:
\begin{equation}
\{A, K\} = 0  ~~ \textbf{and} ~~[A, H]=0
\end{equation}

                  \textbf{It is our definition of symmetry we are looked
                   for.}

             It is interesting that after this operator is
             constucted, we will be able to define relativistic
             supercharges as follows:
$$Q_1=A, ~~~ Q_2 = \imath \frac{AK}{|\kappa|}$$

It is obvious that

$$\{Q_1,Q_2\} = 0, ~~~~ Q_1^2=Q_2^2 $$
 and we can find Witten's superalgebra~\cite{Witten}, where $Q_1^2 \equiv h$ is
a so-called, Witten's Hamiltonian.

  Now our goal is a construction of the operator $A$. We know, that there is a Dirac's $\gamma^5$ matrix,
   that anticommutes with $K$. \textbf{What else?} There is a \textbf{simple theorem}
    ~\cite{Khachidze-2}, according to which
    arbitrary $(\vec{\Sigma} \cdot \vec{V})$
\textbf{type operator}, \textbf{where} $\vec{V}$    \textbf{is a
vector with respect to} $\vec{l}$ \textbf{and is perpendicular to
it}, $(\vec{l}\cdot \vec{V})=(\vec{V} \cdot \vec{l})$,
\textbf{anticommutes with} $K$:

\begin{equation}
\{(\vec{\Sigma}\cdot \vec{V}), K\} = 0
\end{equation}
It is evident that the class of operators anticommuting with $K$
(so-called $K$-odd operators) is much wider.
 Any operator of the form    $\hat{O}(\vec{\Sigma}\cdot \vec{V})$ ,
 \textbf{where $\hat{O}$
commutes with $K$, but is otherwise arbitrary, also is a $K$-odd.}
This fact will be used below.

  Now one can proceed to the second stage of our problem: we wish to construct the $K$-odd operator $A$, that
  commutes with $H$. It is clear that there remains large freedom according to the above mentioned remark about
   $\hat O$ operators - one can take $\hat O$ into account or ignore it.

   We have the following physically interesting vectors at hand which obey the requirements of our theorem. They are
     \begin{equation}   \hat{\vec{r}}  - \textbf{unit radius-vector  and}   ~~   \vec{p}     - \textbf{linear momentum
     vector} \end{equation}
Both of them are perpendicular to $\vec{l}$  ~\cite{footnote}.
Thus, we choose the following $K$-odd terms
\begin{equation} (\vec{\Sigma}\cdot \hat{\vec{r}}), ~~~
K(\vec{\Sigma}\cdot \vec{p}) ~~~ \textbf{and} ~~~ K \gamma^5
\end{equation}
and let probe the combination
\begin{equation} A=x_1(\vec{\Sigma}\cdot \hat{\vec{r}})+\imath x_2
K(\vec{\Sigma}\cdot \vec{p}) + \imath x_3 K \gamma^5 f(r)
\end{equation}

Here the coefficients   $x_i(i=1,2,3)$ are chosen in such a way
that $A$ operator is Hermitian for arbitrary real numbers and
$f(r)$ is an arbitrary scalar function to be determined later from
the symmetry requirements. Let's calculate
\begin{eqnarray}
0=[A,H]=(\vec{\Sigma}\cdot \hat{\vec{r}})\{x_2 V'(r)-x_3 f'(r)\}
+\\ +\nonumber{2 \imath \beta K \gamma^5 \{\frac{x_1}{r} - m x_3
f(r)\}}
\end{eqnarray}

  We have a diagonal matrix in the first row, while the antidiagonal matrix in the second one.
  Therefore two equations follow:
  \begin{equation}
  x_2 V'(r) = x_3 f'(r),~~~~
  x_3 m f(r)=\frac{x_1}{r}
  \end{equation}

One can find from these equations:
\begin{equation}
V(r)= \frac{x_1}{x_2} \frac{1}{mr}
\end{equation}
\textbf{Therefore only the Coulomb potential corresponds to the
above required $\pm \kappa$ degeneracy.}
   The final form of obtained $A$ operator is the following:
\begin{equation}
A=x_1 \{(\vec{\Sigma}\cdot \hat{\vec{r}})-
\frac{\imath}{ma}K(\vec{\Sigma}\cdot \vec{p}) + \frac{\imath}{mr}K
\gamma^5\}
\end{equation}
where unessential common factor $x_1$ may be omitted and after
using known relations for Dirac matrices, this expresion may be
reduced to the form
\begin{equation}
A = \gamma^5 \{ \vec{\alpha} \cdot\hat{ \vec{r}}  -
\frac{\imath}{m a}K \gamma^5(H- \beta m)\}
\end{equation}

Above and here $a$ is a strengh of Coulomb potential, $a=Z
\alpha$. Precisely this operator is given in Johnson's and
Lippmann's abstract~\cite{Johnson}.

\textbf{What the real physical picture is standing behind this?}
Taking into account the relation

\begin{equation}
K \left(\vec{\Sigma}\cdot\vec{p}\right)=
 - \imath \beta \left( \vec{\Sigma} \cdot
\frac{1}{2}[\vec{p} \times \vec{l} - \vec{l} \times
\vec{p}]\right)
\end{equation}
one can recast our operator in the following form $$ A =
\vec{\Sigma}\cdot \left(\hat{\vec{r}} - \frac{\imath}{2ma} \beta
\left[\vec{p}\times\vec{l}-\vec{l}\times\vec{p}\right]\right)+\frac{\imath}{mr}K
\gamma^5 $$ One can see that in the non-relativistic limit, i.e.
$\beta\rightarrow 1$ and $\gamma^5 \rightarrow 0$ our operator
reduces to
\begin{equation}
A \rightarrow A_{NR}=
\vec{\sigma}\cdot\left(\hat{\vec{r}}-\frac{\imath}{2ma}[\vec{p}\times\vec{l}-
\vec{l}\times\vec{p}]\right)
\end{equation}

Note the LRL vector in the parenthesis of this equation.
\textbf{Therefore relativistic supercharge reduces to the
 projection of the LRL vector on the electron spin direction. Precisely this operator was used in the case of Pauli
electron~\cite {Tangerman}.}

  Because the Witten's Hamiltonian is
  \begin{equation}
  A^2=1+\left(\frac{K}{a}\right)^2 \left(\frac{H^2}{m^2}-1\right)
  \end{equation}
and it consists only mutually commuting operators, it is possible
their simultaneous diagonalization and replacement by
corresponding eigenvalues. For instance, the energy of ground
state is
\begin{equation}
E_0=\left(1-\frac{(Z\alpha)^2}{\kappa^2}\right)^{1/2}
\end{equation}
By using the ladder procedure, familiar for SUSY quantum
mechanics, the Sommerfeld formula can be easily
 derived~\cite{Katsura}.

  Let's remark that if we include the Lorentz-scalar potential as well
\begin{equation}
H=\vec{\alpha}\cdot \vec{p}+\beta m +V(r) + \beta S(r)
\end{equation}
this last Hamiltonian also commutes with $K$-operator, but
\textit{does not commute} with the above JL operator.

  On the other hand, non-relativistic quantum mechanics is \textbf{indifferent} with regard
  of \textbf{the Lorentz variance
   properties} of potential. Therefore it is expected that in case of scalar potential the description of
    hidden symmetry should also be possible. In other words, \textbf{the JL operator must be generalised} to this case.

   For this purpose it is necessary to increase number of $K$-odd structures. One has to use our theorem
   in the part of additional $\hat{O}$ factors.

  Now let us probe the following operator
  \begin{equation}
  B=x_1\left(\vec{\Sigma}\cdot\hat{\vec{r}}\right)+x_1'\left(\vec{\Sigma}\cdot\hat{\vec{r}}\right)H
  +\imath x_2 K \left(\vec{\Sigma}\cdot \vec{p} \right) + \imath
  x_3 K \gamma^5 f_1(r)+\imath x_4 K \gamma^5 \beta f_2(r)
  \end{equation}

We included $\hat{O}=H$ in the first structure and at the same
time – the matrix $\hat{O}=\beta$ in the third structure.
 Both of them commute with $K$ and are permissible by our theorem. The form (20) is a minimal extention of the
  previous picture, because only the first order structures in $\hat{\vec {r}}$ and $\vec{p}$ participate. For turning to the previous
   case one must take $x_3=x_4=0$ and $S(r)$=0.
Calculation of relevant commutators give:
\begin{eqnarray}
[B,H] = \gamma^5 \beta K \{\frac{2\imath x_1}{r}- 2 \imath x_3
(m+S) f_1(r)+\frac{2\imath x_1'}{r}V(r)\}+\\ \nonumber{ + K
(\vec{\Sigma}\cdot \hat{\vec{r}}) \{x_2 V'(r) - x_3 f_1'(r)\}}+\\
\nonumber{
 + K\beta \left( \vec{\Sigma}\cdot \hat{\vec{r}}\right) \{x_2 S'(r)-x_4 f_2'(r)\}+}
 \\
\nonumber{+ \gamma^5 K \{ \frac{2\imath x_1'(m+S)}{r}-2\imath x_4
(m+S)f_2(r)\}+}\\ \nonumber{ + \beta K \{ \frac{2\imath x_1'}{r}-2
\imath x_4 f_2(r)\} (\vec{\Sigma} \cdot \vec{p})}
\end{eqnarray}
Equating this expression to zero,we derive matrix equation, then
passing to $2\times2$ representation we must equate to zero the
coefficients standing in fronts of diagonal and antidiagonal
elements. In this way it follows equations:

   \textbf{(i) From antidiagonal structures} ($\gamma^5,~ \gamma^5 \beta K$)

\begin{eqnarray}
~~\frac{x_1}{r}-x_3(m+S)f_1(r)+\frac{x_1'}{r}V(r)=0 \\
\nonumber{\frac{x_1'}{r}(m+S)-(m+S)f_2(r)=0}
\end{eqnarray}

   \textbf{(ii) From diagonal structures}~( $K(\vec{\Sigma} \cdot \hat{\vec{r}}),~~K \beta
    (\vec{\Sigma}\cdot \hat{\vec{r}}), ~~ \beta K (\vec{\Sigma}\cdot \vec{p}) ~ $):
    \begin{eqnarray}
    x_2 V'(r)-x_3f_1'(r)=0\\
    \nonumber{\frac{x_1}{r}-x_2(m+S)V(r)+\frac{x_1'}{r}V(r)=0}\\
    \nonumber{x_2 S'(r)-x_4f_2'(r)=0}\\
    \nonumber{\frac{x_1'}{r}-x_4 f_2(r)=0}
\end{eqnarray}

  Integrating the first and third equations in (23) for vanishing boundary conditions at infinity, we obtain
\begin{eqnarray}
f_1(r)=\frac{x_2}{x_3}V(r), ~~~~ f_2(r)=\frac{x_2}{x_4}S(r)
\end{eqnarray}
and taking into account the last equation from (23), we have
\begin{equation}
f_2(r)=\frac{x_1'}{x_4 r}
\end{equation}
Therefore, according to (24), we obtain finally
\begin{equation}
S(r)=\frac{x_1'}{x_2 r}
\end{equation}

\textbf{So, the scalar potential must be Coulombic.}

 Inserting
Eq.(24) into the first equation of (22) and solving for $V(r)$,
one derives
\begin{equation}
V(r)=\frac{x_1}{r} \frac{1}{x_2(m+S)-{\frac{x_1'}{r}}}
\end{equation}
At last, using here the expression (26), we find
\begin{equation}
V(r) = \frac{x_1}{x_2 m r}
\end{equation}

 \textbf{Therefore we have asserted that the $\pm \kappa$ degeneracy is a symmetry of the Dirac equation only for Coulomb
 potential
 (for any general combination of Lorentz scalar and 4th component of a Lorentz
 vector).}

  \textbf{ Proceed from this fact, we believe that the nature of "accidental" or "hidden" symmetry of the Kepler problem
  is
   established - it is a $\pm \kappa$ degeneracy. Generators of this symmetry, $A$ or $B$ describe this degeneracy - they
   interchange these two values.}

 It is evident from an ordinary momentum algebra that the label  $\kappa=j+1/2$ takes place for aligned spin $j=l+1/2$,
 i.e.
  for states ($s_{1/2},p_{3/2},$ etc.), while $\kappa=-(j+1/2)$ holds for unaligned
  spin or for states ($p_{1/2},d_{3/2},$ etc.).
  \textbf{We see, that
   the  $\pm \kappa$ degeneracy leads to the forbidden of the Lamb shift. Therefore absence of the Lamb shift in the
   Dirac equation for pure Coulomb potential is a consequence of the above mentioned symmetry, which in its deep meaning is
   related to the conservation
    of LRL vector (!).}

 \textbf{After all this by means of known ways the $SO(4)$ algebra could be constructed, manifestation of which in our
 macroworld is a conservation of the LRL vector and cosequently closeness of celestial
 orbits.}

  Therefore, it seems that the conservation of LRL vector is a macroscopic manifestation of symmetry, which is present
  in microworld.

  Further, if we take into account above obtained solutions, one can reduce the $B$ operator to more compact form
  \begin{equation}
  B = (\vec{\Sigma}\cdot\hat{\vec{r}})(ma_V+a_SH)-\imath K
  \gamma^5(H-\beta m)
  \end{equation}
where the following notations are used $$ a_V= - \frac{x_1}{x_2
m}, ~~~ ~~~ a_S= - \frac{x_1'}{x_2} $$
 Here $a_i$-s are the
constants of corresponding Coulomb potentials

$$V(r)=-\frac{a_V}{r},~~~~~~~S(r)=-\frac{a_S}{r} $$

 In conclusion we want to remark, that this expression for $B$ was mentioned recently by Leviatan~\cite{Leviatan}, who
 used the radial decomposition and separation of spherical angles in the Dirac equation.

   \textbf{Our approach is 3-dimensional, without any referring to radial equation and, therefore is more general and rather easy.
    Moreover we find the source of mysterious "hidden" symmetry of the Kepler problem, as the degeneracy on quantum
    level.}

}


\begin{thebibliography}{99}
\bibitem{footnote}
    Constraints of this theorem are also satisfied by
Laplace-Runge-Lenz vector $\vec{A}=\hat
{\vec{r}}-\frac{i}{2ma}[\vec{p}\times \vec{l}-\vec{l}\times
\vec{p}]$, but this vector is associated to the Coulomb potential.
Hence, we abstain from its consideration for now. Moreover
$(\vec{\Sigma}\cdot\vec{A})$  is not an independent structure. It
is expressible by two other structures, e.g.\
$(\vec{\Sigma}\cdot\vec{A})=(\vec{\Sigma}\cdot\hat{\vec{r}})+\frac{i}{ma}\beta
K(\vec{\Sigma}\cdot\vec{p})$\ ;
\end{thebibliography}
\end{document}